\documentclass[12pt]{article}
\usepackage[english]{babel}
\usepackage{amssymb,amsmath}
\usepackage{graphicx}
\usepackage{multicol}
\usepackage{multirow}
\usepackage{slashbox}
\usepackage{subfigure}
\usepackage{times}

\usepackage{epsf}
\usepackage{hyperref}
\hoffset= -1.5cm \voffset= -2cm \vspace{2cm}

\newcommand{\be}{\begin{equation}}
\newcommand{\ee}{\end{equation}}

\begin{document}
\title{ Dispersive Approach to Abelian Axial Anomaly, Mixing of Pseudoscalar Mesons and Symmetries }
\date{}
\author{Y.~N.~Klopot$^1$\footnote{{\bf e-mail}: klopot@theor.jinr.ru}\;,\,
        A.~G.~Oganesian$^2$\footnote{{\bf e-mail}: armen@itep.ru}\;\, and \
        O.~V.~Teryaev$^1$\footnote{{\bf e-mail}: teryaev@theor.jinr.ru}}
\maketitle
 \hspace{-7mm}
  {$^{1}$\em Bogoliubov Laboratory of Theoretical Physics, Joint Institute for Nuclear Research,\;\; Dubna, Russia, 141980.\\
  $^{2}$Institute of Theoretical and Experimental Physics,\;\;B.Cheremushkinskaya 25, Moscow, Russia, 117218\\
      }
\begin{abstract}

We suggest a rigorous generalization of the pseudoscalar mesons
mixing description in $SU(3)$ basis. It is shown that the
appearance of extra massive state nicknamed in the paper as
glueball is unavoidable in any scheme with more than one angle.
In this framework we develop the dispersive approach to Abelian
axial anomaly of isoscalar non-singlet current. Combining it with
the analysis of experimental data of charmonium radiative decays
ratio we get the number of quite strict constraints for mixing
parameters. Our analysis favors the equal values of axial
currents coupling constants which may be considered
as a manifestation of
$SU(3)$ symmetry and possible
violation of chiral symmetry based
predictions.

\end{abstract}

\section{Introduction}

The problem of pseudoscalar meson states mixing has been under
intense theoretical and experimental study for many years. This
topic attracts a lot of interest due to its close relation to
such fundamental phenomena  as quantum anomaly, chiral symmetry
breaking and study of exotic states like glueball, which can
elucidate the essential features of QCD.

During the last decade, a large amount of new experimental data
on mesons has been collected. New data from the forthcoming
experiments at COMPASS, BES-III, GlueX (JLAB) and at the upgraded
facilities FAIR (GSI) and MAMI  might allow a complete
quantitative verification of the different mixing schemes and
approaches.

A number of analyzes of mixing in $\eta-\eta'$ system based on different processes have been
performed in the last decades and the mixing angles in the range
$-10^\circ \div-24^\circ$ were obtained (see, for example
\cite{Diakonov:1981nv} -- \cite{Bhagwat:2007ha} ). The analysis
of the axial anomaly generated decays $\eta(\eta')\to2\gamma$ was
also performed in \cite{Donoghue:1986wv} (in the framework of
ChPT) and \cite{Gilman:1987ax}, and the estimation $\theta$ =
$-20^\circ\div 25^\circ$ was obtained.

The approach with one mixing angle dominated for decades in the
studies of the $\eta$-$\eta'$ system. However, in the recent
years there was a  rise of interest in the mixing schemes with
two and three angles. The theoretical ground of this was based on
the observation that taking into account the chiral anomaly
through perturbative expansion in ChPT  can lead to the
introduction of two mixing angles in the description of the
$\eta$-$\eta'$ system \cite{Leutwyler:1997yr,Kaiser:1998ds}. Some
analyzes of various decay processes were also performed  in this
scheme
\cite{Feldmann:1998vh,Feldmann:1998sh,DeFazio:2000my,Kroll:2005sd,Escribano:2005qq}.

The theoretical analysis of the mixing in the pseudoscalar
sector  base either on  $SU(3)$ or quark basis. The last was
introduced by T.~Feldmann and P.~Kroll \cite{Feldmann:1998vh} and
was widely used in the last years.

For us it happens to be more convenient to construct and use the
rigorous generalization of $SU(3)$ basis similar to the mixing of
massive neutrinos. This is because we use the dispersive approach
to axial anomaly \cite{Dolgov:1971ri} (see also \cite
{Ioffe:2006ww} for the review) to find some model-independent and
precise restriction on the mixing parameters (angles and decay
constants) in the different schemes (with one and more angles).
The combination of our approach with certain processes leads to
quite strict predictions. We consider this paper as a first step
and application of our approach to the whole set of processes is
to be done.

This paper is organized in the following way. In Section 2 we
introduce the general mixing scheme in $SU(3)$ basis. More
specifically, we derive the expression for (non-square) matrix of
coupling constants generalizing the approach offered by B.~Ioffe
and M.~Shifman \cite{Ioffe:1979rv,Ioffe:1980mx}. In Section 3 we
consider the dispersive approach to axial anomaly for the
isoscalar non-singlet axial current $J_{\mu5}^8$ taking into
account the possible contributions of higher mass states. In this
way we derive our main equation while in Section 4 we supplement
it by the analysis of the ratio of radiative decays of $J/\Psi$.
These two constraints happen to be sufficient to provide strong
bounds for mixing parameters. In Section 5 we consider the
reduction of the general scheme with three angles to some
currently popular particular (two-angle) schemes
\cite{Kou:1999tt} -- \cite{Cheng:2008ss}. Finally, in  Section 6
we summarize the results of numerical analysis and discuss the
possible implications  for $SU(3)$ and chiral symmetries.

\section{Mixing}
We start with a vector of physical pseudoscalar fields consisting
of the fields of lightest pseudoscalar mesons and other fields:
\be \widetilde{\varPhi}\equiv
\begin{pmatrix}
\pi^0\\
\eta\\
\eta'\\
G \\
\vdots
\end{pmatrix}
\ee We do not need to specify the physical nature of other
components with higher masses, the lower of which $G$ is probably
a glueball. Let us also introduce, following
\cite{Ioffe:1979rv,Ioffe:1980mx}, a set of $SU(3)$ fields
$\varphi_k (k=3,8,0)$ and other (singlet) fields $g_i$: \be
\label{Phi} \mathbf{\varPhi}=
\begin{pmatrix}
\varphi_3\\
\varphi_8\\
\varphi_0\\
g\\
\vdots
\end{pmatrix},\;\;\;
\ee
and corresponding states $|P_k \rangle$ and $|P_{k'} \rangle$.
The three upper fields $\varphi_3, \varphi_8, \varphi_0$
diagonalize the matrix elements of axial currents
$J_{\mu5}^l=\overline{q}\gamma_\mu\gamma_5\frac{\lambda^l}{\sqrt{2}}q$
:

\be \label{DecConst} \langle 0 \vert J^{l}_{\mu 5}\vert P_k
\rangle = i \delta_{lk} f_k q_{\mu}\;, ~~ l,~k = 3,8,0.\, \ee All
other states are orthogonal to these currents: \be \langle 0
\vert J^{l}_{\mu 5}\vert P_{k'} \rangle = 0. \ee
At the same time all the corresponding fields enter the mass term
in the effective Lagrangian with a generally speaking
non-diagonal mass matrix
  $\mathbf{M}$ :

   \be\Delta \mathcal{L}= \frac{1}{2} \mathbf{\varPhi^T M \varPhi}
\ee This formula immediately implies the generalized PCAC
relation:

\be \label{J_diverg1} \mathbf{\partial_\mu J_{\mu5} = FM\varPhi},
\ee where

\be \mathbf{\partial_\mu J_{\mu5}}\equiv
\begin{pmatrix}
\partial_\mu J_{\mu5}^3\\
\partial_\mu J_{\mu5}^8\\
\partial_\mu J_{\mu5}^0
\end{pmatrix},\;
\mathbf{F}\equiv \begin{pmatrix}
f_3 & 0 & 0 & 0 & \ldots & 0\\
0 & f_8 & 0 & 0 & \ldots & 0\\
0 & 0 & f_0 & 0 & \ldots & 0\\
\end{pmatrix},
\ee $\mathbf{F}$ is a matrix of decay constants\footnote{Note,
that matrix of decay constants $\mathbf{F}$ is  non-square
expressing the fact that generally the number of $SU(3)$ currents
is less then the number of all possible states involved in
mixing.   The similar situation takes place (see e.g.
\cite{Bilenky:1998dt}) in one of the extensions of the Standard
Model --- neutrino mixing scenario involving sterile neutrinos.}
defined in (\ref{DecConst}).

 In order to proceed from initial $SU(3)$ fields $\mathbf{\varPhi}$ to physical mass fields $\widetilde{\varPhi}$ the unitary matrix U is introduced

\be \mathbf{\widetilde{\varPhi}=U\varPhi} \ee which diagonalizes
the mass matrix \be \mathbf{UMU^{T}=\widetilde{M}}\equiv
diag(m_{\pi^0}^2,m_{\eta}^2,m_{\eta'}^2,m_G^2, \ldots), \ee
where $m_\pi$, $m_{\eta}$, $m_{\eta'}$ and $m_{G}$ are the masses
of the $\pi$, $\eta$, $\eta'$ mesons and glueall state $G$
respectively.

Simple transformations of Eq.(\ref{J_diverg1}) reads: \be
\mathbf{\partial_\mu J_{\mu5} =
FU^{T}\widetilde{M}\widetilde{\varPhi}} \ee
This formula is close to those obtained in
\cite{Ioffe:1979rv,Ioffe:1980mx} (in the limit of small mixing).
When the decay constants are equal this is reduced to formula
(3.40) in \cite{Diakonov:1995qy}.

Taking into account the well-known smallness of $\pi^0$ mixing
with $\eta, \eta'$ sector
\cite{Ioffe:1979rv,Ioffe:1980mx,Ioffe:2007eg} and neglecting all
higher contributions we restrict our consideration to three
physical states $\eta, \eta', G$ and two currents $J_{\mu5}^8,
J_{\mu5}^0 $.  Then the divergencies of the axial currents
(recall, that $G$ is a first mass state heavier than $\eta'$):
\be \label{DivJ_FUMPhi}
\begin{pmatrix}
\partial_\mu J_{\mu5}^8\\
\partial_\mu J_{\mu5}^0\\
\end{pmatrix}
=
\begin{pmatrix}
f_8 & 0 & 0\\
0 & f_0 & 0\\
\end{pmatrix}
\mathbf{U^{T}}
\begin{pmatrix}
m_{\eta}^2 & 0 & 0\\
0 & m_{\eta'}^2 & 0\\
0 & 0 & m_{G}^2
\end{pmatrix}
\begin{pmatrix}
\eta\\
\eta'\\
G
\end{pmatrix}
\ee

Exploring the mentioned similarity of meson and lepton mixing,
we use a well-known general Euler parametrization for the mixing
matrix\footnote{We use notation $c_i\equiv cos\theta_i, s_i\equiv
sin\theta_i$} $\mathbf{U}$: \be \label{MixMat}
\mathbf{U}=\begin{pmatrix}
c_3 & -s_3 & 0\\
s_3 & c_3 & 0\\
0 & 0 & 1
\end{pmatrix}
\begin{pmatrix}
1 & 0 & 0\\
0 & c_0 & -s_0\\
0 & s_0 & c_0
\end{pmatrix}
\begin{pmatrix}
c_8 & -s_8 & 0\\
s_8 & c_8 & 0\\
0 & 0 & 1
\end{pmatrix}
=
\begin{pmatrix}
c_8c_3-c_0s_3s_8 & -c_3s_8-c_8c_0s_3 & s_3s_0\\
s_3c_8+c_3c_0s_8 & -s_3s_8+c_3c_8c_0 & -c_3s_0\\
s_8s_0 & c_8s_0 & c_0
\end{pmatrix}
\ee

As soon as in the chiral limit $J_{\mu5}^8$ should be conserved,
Eq.(\ref{DivJ_FUMPhi}) obviously implies, that the
coefficients in front of the terms $m_{\eta}^2, m_{\eta'}^2,
m_G^2$ should decrease at least as $(m_\eta/m_{\eta',G})^2$.

The generic matrix (\ref{MixMat}) can be reduced to different
particular cases. The pure $\eta-\eta'$ mixing with no other
admixtures corresponds to the case $\theta_0=0$ in Eq.
(\ref{MixMat}). This is the so-called one-angle mixing scheme
with the mixing angle $\theta=\theta_8+\theta_3$. As it was
mentioned above, that this scheme is not sufficient for
description of the full set of experiments.

One can easily see from (\ref{DivJ_FUMPhi}), (\ref{MixMat}) that
the schemes with more than one mixing parameter unavoidably
require introduction of new states for the mixing matrix to be
unitary. Indeed, from these equations one can see that in general
case we have 3 different angles and in some particular cases one
can reduce the number of angles by one and to get the schemes
with two different angles (put $\theta_3=0$ or  $\theta_8=0$).
Subsequently  the schemes with 3 and 2 angles will be considered
in details in sections 4 and 5.

In this work we sequentially use $SU(3)$ basis. For our purposes
this basis is more preferable since in the next Section we will
consider non-singlet isoscalar axial current $J_{\mu5}^8$ which
is free from non-Abelian anomaly. The transition to quark basis
$\mathbf{\Phi_q}=((u\bar{u}+d\bar{d})/\sqrt{2}, s\bar{s}, g)$
which is also widely used in literature can be performed by
multiplying by additional rotation matrix $\mathbf{V}$: \be
\label{qbasis} \mathbf {\Phi=V\Phi_q} ,\;\; \mathbf{V}=
\begin{pmatrix}
\sqrt{1/3} & -\sqrt{2/3} & 0\\
\sqrt{2/3} & \sqrt{1/3} & 0\\
0 & 0 & 1
\end{pmatrix}.
\ee

\section{Dispersive approach to axial anomaly}
In our paper the dispersive form of the anomaly sum rule will be
extensively used, so we remind briefly the main points of this
approach (see e.g. review \cite{Ioffe:2006ww} for details).

Consider a  matrix element of a transition of the axial current
to two photons with momenta $p$ and $p'$ \be
 T_{\mu \alpha \beta} (p, p') = \langle p, p' \vert J_{\mu 5}
\vert 0 \rangle \;. \ee

The general form  of $T_{\mu \alpha \beta}$ for a case $p^2=p'^2$
can be represented in terms of structure functions (form factors):

\be T_{\mu \alpha \beta}(p, p') = F_1(q^2) q_{\mu}
\epsilon_{\alpha \beta \rho \sigma} p_{\rho} p'_{\sigma} +
\frac{1}{2} F_2 (q^2) [\frac{p_{\alpha}}{p^2} \epsilon_{\mu \beta
\rho \sigma}p_{\rho}p'_{\sigma} - \frac{p'_{\beta}}{p^2}
\epsilon_{\mu \alpha \rho
\sigma}p_{\rho}p'_{\sigma}-\epsilon_{\mu \alpha \beta
\sigma}(p-p')_{\sigma}] \;, \ee where $q=p+p'$. The functions
$F_1(q^2)$, $F_2(q^2)$ can be described by dispersion relations
with no subtractions and anomaly condition in QCD results in a
sum rule:

\be \label{sumrule} \int\limits^{\infty}_{0}~ Im~F_1(q^2) dq^2 =
2\alpha N_c\sum e_q^2\;, \ee where $e_q$ are quark electric
charges and $N_c$ is the number of colors. This sum rule was
proved  \cite{Veretin:1994dn} for the general case
$p^2\not=p^{\prime 2}$ and earlier in \cite{Frishman:1980dq} and
\cite{Horejsi:1985qu} for the cases $p^2 <0, m=0$ and
$p^2=p^{\prime2}$ respectively. Notice that in QCD this equation
does not have any perturbative corrections
\cite{Adler:1969er}, and it is expected that it does not have any
non-perturbative corrections also due to 't Hooft's consistency
principle \cite{Horejsi:1994aj}. It will be important for us that
at $q^2 \to \infty$ the function $ImF_1(q^2)$ decreases as
$1/q^4$. Note also that the relation (\ref{sumrule}) contains
only mass-independent terms, which is especially important for
the 8th component of the axial current $J_{\mu5}^8$ containing
strange quarks:

\be
 J^{(8)}_{\mu 5} = \frac{1}{\sqrt{6}}(\bar{u} \gamma_{\mu} \gamma_5 u + \bar{d}
\gamma_{\mu} \gamma_5 d - 2\bar{s} \gamma_{\mu} \gamma_5 s)  \;
\ee

The general sum rule (\ref{sumrule}) takes the form:

\be \label{sumrule8} \int\limits^{\infty}_{0}~ Im~F_1(q^2) dq^2 =
\frac{2}{\sqrt{6}}\alpha (e^2_u + e^2_d - 2e^2_s) N_c =
\sqrt{\frac{2}{3}}\alpha  \;, \ee
 where $e_u=2/3$,\; $e_d=e_s=-1/3$, $N_c=3$.

Consider now a particular case of pure $\eta-\eta'$ mixing, where no other mixing states are taken into account.
Recall that this case has been studied for a long time, various approaches were considered and
the numbers for the mixing angles in the range $-(10\div24)$ were obtained.
The approach based on the dispersive representation of axial
anomaly was introduced for $\eta-\pi^0$ mixing in
\cite{Ioffe:2007eg} and used in \cite{Klopot:2008ec} where
$\eta-\eta'$ mixing was considered in assumption of small mixing
angle).

In order to separate the form factor $F_1(q^2)$, multiply $T_{\mu
\alpha \beta} (p, p')$ by $q_\mu /q^2$. Then taking the imaginary
part of $F_1(q^2)$, using the expression for $\partial_\mu
J_{\mu5}^8$ from Eq.(\ref{DivJ_FUMPhi}) and saturating the matrix
element with the $\eta, \eta'$ states we get:

%

$$Im F_1(q^2)=Im~ q_{\mu}\frac{1}{q^2} \langle 2\gamma \mid J^{(8)}_{\mu 5}\mid0 \rangle=$$
$$ -\frac{f_8}{q^2}\langle 2\gamma \mid[ m_{\eta}^2\eta(c_1) + m_{\eta'}^2\eta'(s_1)]\mid0 \rangle=$$
\be \pi
f_8[A_{\eta}\delta(q^2-m_{\eta}^2)(c_1)+A_{\eta'}\delta(q^2-m_{\eta'}^2)(s_1)]
\ee

Note, that if we have included higher resonances to this
equation, they are expected to be suppressed as $1/m^4_{res}$ by
virtue of the mentioned above asymptotical behavior of
$F_1(q^2)\propto 1/q^4$. This (approximate) independence on
higher resonances together with (exact) quark mass independence
of the anomaly relation (\ref{sumrule}) may be an indication of
some connection between these two effects.

If we employ the sum rule (\ref{sumrule8}), we obtain a simple
equation:

\be \label{J8_anom1} c_1+\beta s_1=\xi, \ee where \be
\beta\equiv\frac{A_{\eta'}}{A_{\eta}}=\sqrt{\frac{\Gamma_{\eta'\to
2\gamma}}{\Gamma_{\eta \to
2\gamma}}\frac{m_{\eta}^3}{m_{\eta'}^3}}, \;\;\;
\ee \be \xi\equiv\sqrt{\frac{\alpha^2
m^3_{\eta}}{96{\pi}^3\Gamma_{{\eta} \to 2\gamma}}\frac{1}{f^2_8}}
,\;\;\;\;
\Gamma_{\eta\to2\gamma}=\frac{m_{\eta}^3}{64\pi}A_\eta^2\;. \ee


\begin{figure}
\centerline{\includegraphics[width=0.5\textwidth]{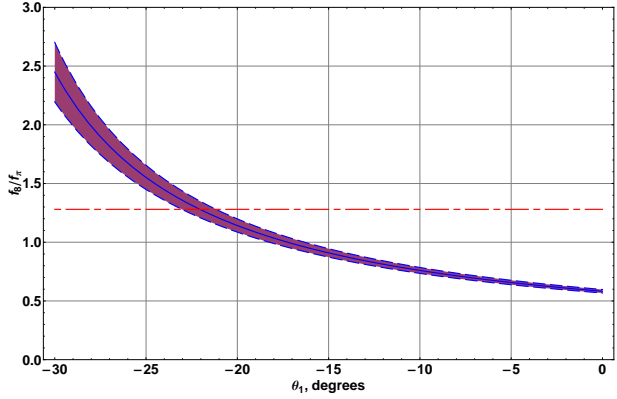}}
\caption{Mixing angle $\theta_1$ as a function of the decay
constant $f_8$ in the one-angle mixing scheme. Dashed curves
correspond to the uncertainties of the experimental data input.
Horizontal dot-dashed line indicates the  $f_8=1.28f_\pi$ level.}
\label{1ang-plot1}
\end{figure}

For numerical evaluation of the $\eta-\eta'$ mixing angle in the
latter  case, put $f_8=1.28f_\pi$ , $f_\pi=130.4$ MeV,
$m_\eta=547.85$ MeV, $m_\eta'=957.78$ MeV, $\Gamma_{\eta \to
2\gamma}=0.51$ keV,  $\Gamma_{\eta' \to 2\gamma}=4.30 $ keV. The
value for $f_8$  is taken from the  chiral perturbation theory
calculations, other numbers from PDG Review
2008 \cite{Amsler:2008zzb}. The mixing angle, responsible for the
mixing of the $\eta-\eta'$ system appears to be $\theta_1 =
-22.1^\circ \pm 1.5^\circ$. The dependence of the mixing angle on
$f_8$ is shown on the Fig.\ref{1ang-plot1}. Remarkably, the
anomaly sum rule \textit{fixes} the mixing angle (provided we
know $f_8$) in this case.

From our point of view one of the advantages of our approach is a
high accuracy due to a high accuracy of the anomalous sum rule
(\ref{sumrule8}) for octet axial current. Let us stress that at
this stage we avoid consideration of the anomaly relation for the
singlet axial current which contains the contributions of gluons,
direct instantons and topological effects (see e.g.
\cite{Geshkenbein:1979vb}, \cite{'tHooft:1999jc}). As a result it
appears to be unnecessary to include $f_0$ to our analysis.

The current result moderately agrees with our earlier analysis in
the small mixing angle approximation \cite{Klopot:2008ec}. It is
not completely trivial that our result is also in a good
agreement with previous analysis done in pioneering papers
\cite{Akhoury:1987ed,Ball:1995zv}. This is because
Eq.(\ref{J8_anom1}) happens to follow also from the
(non-dispersive) anomaly equations used in
\cite{Akhoury:1987ed,Ball:1995zv}. The similar result of
dispersive and non-dispersive (local) approaches is in some sense
natural taking into account that we omit the higher contributions
with controlled accuracy $\mathcal{O}(1/m^4)$.
At the same time the mentioned approaches actually use the anomalous divergency for singlet current which we do not need.

Now let us consider a generic case involving glueball admixtures.
Performing the same operations for the $\eta-\eta'-G$ system we
get:

$$Im F_1(q^2)=Im~ q_{\mu}\frac{1}{q^2} \langle 2\gamma \mid J^{(8)}_{\mu 5}\mid0 \rangle=$$
$$ -\frac{f_8}{q^2}\langle 2\gamma \mid[ m_{\eta}^2\eta(c_8c_3-c_0s_3s_8) + m_{\eta'}^2\eta'(s_3c_8+c_3c_0s_8) + m_{G}^2 G s_8s_0 ]\mid0 \rangle=$$
\be \pi
f_8[A_{\eta}\delta(q^2-m_{\eta}^2)(c_8c_3-c_0s_3s_8)+A_{\eta'}\delta(q^2-m_{\eta'}^2)(s_3c_8+c_3c_0s_8)+A_{G}\delta(q^2-m_{G}^2)(s_8s_0)]
\ee The final equation following from the anomaly sum rule
(\ref{sumrule8}) for the $\eta-\eta'-G$ system is:

\be \label{J8_anom_final}
(c_8c_3-c_0s_3s_8)+\beta(s_3c_8+c_3c_0s_8)+\gamma(s_8s_0)=\xi,
\ee where \be
\beta\equiv\frac{A_{\eta'}}{A_{\eta}}=\sqrt{\frac{\Gamma_{\eta'\to
2\gamma}}{\Gamma_{\eta \to
2\gamma}}\frac{m_{\eta}^3}{m_{\eta'}^3}}, \;\;\;
\gamma\equiv\frac{A_{G}}{A_{\eta}}=\sqrt{\frac{\Gamma_{G\to
2\gamma}}{\Gamma_{\eta \to 2\gamma}}\frac{m_{\eta}^3}{m_{G}^3}};
\ee \be \xi\equiv\sqrt{\frac{\alpha^2
m^3_{\eta}}{96{\pi}^3\Gamma_{{\eta} \to 2\gamma}}\frac{1}{f^2_8}}
,\;\;\;\;
\Gamma_{\eta\to2\gamma}=\frac{m_{\eta}^3}{64\pi}A_\eta^2\;. \ee

Let us summarize the situation with a  theoretical accuracy of the Eq.
(\ref{J8_anom_final}).  As we have pointed out, the anomaly sum
rule (\ref{sumrule8}) has no
$\alpha_s$-corrections\footnote{However, this is not the case for
the singlet axial current $J_{\mu5}^0$.}.
The possible contributions from higher states may come as
the additional terms in the l.h.s. of Eq. (\ref{J8_anom_final}).
As it was discussed
before, the asymptotic behavior  $F_1(q^2)$ is proportional to $1/q^4$
at large $q^2$ and a sort of quark-hadron duality implies that
higher resonances should be suppressed as $(m_{\eta'}/m_{res})^4$.

\section{$J/\Psi$ radiative decay ratio}

Eq. (\ref{J8_anom1}) provide an exact constraint but contains too
many free parameters. As an additional experimental constraint
we, following \cite{Akhoury:1987ed}, \cite{Ball:1995zv} use the
data of decay ratio $R_{J/\Psi}=(\Gamma(J/\Psi)\to
\eta'\gamma)/(\Gamma(J/\Psi)\to \eta\gamma)$.

As it was pointed out in \cite{Novikov:1979uy} the radiative
decays $J/\Psi \to \eta (\eta')\gamma$ are dominated by
non-perturbative gluonic matrix elements, and the ratio of the
decay rates $R_{J/\Psi}=(\Gamma(J/\Psi)\to
\eta'\gamma)/(\Gamma(J/\Psi)\to \eta\gamma)$ can be expressed as
follows: \be \label{RJP1} R_{J/\Psi}=\left|\frac{\langle0\mid
G\widetilde{G}\mid\eta'\rangle}{\langle0\mid
G\widetilde{G}\mid\eta\rangle}\right|^2\left(\frac{p_{\eta'}}{p_{\eta}}\right)^3,
\ee where
$p_{\eta(\eta')}=M_{J/\Psi}(1-m^2_{\eta(\eta')}/M^2_{J/\Psi})/2$.
The advantage of such a choice is an expected smallness of
perturbative and non-perturbative corrections.


The divergencies of singlet and octet components of axial current
in terms of quark fields can be written as:

\be \label{RRPP1}
\partial_\mu J_{\mu5}^8=\frac{1}{\sqrt{6}}(m_u \overline{u}\gamma_\mu\gamma_5u+ m_d \overline{d}\gamma_\mu\gamma_5d- 2m_s\overline{s}\gamma_\mu\gamma_5s ),
\ee \be \label{RRPP2}
\partial_\mu J_{\mu5}^0=\frac{1}{\sqrt{3}}(m_u \overline{u}\gamma_\mu\gamma_5u+ m_d \overline{d}\gamma_\mu\gamma_5d + m_s\overline{s}\gamma_\mu\gamma_5s )+\frac{1}{2\sqrt{3}}\frac{3\alpha_s}{4\pi}G\widetilde{G}
\ee

Following \cite{Akhoury:1987ed}, neglect the contribution of u-
and d- quark masses, then the matrix elements of the anomaly term
between the vacuum an $\eta, \eta'$ states are :
  \be \label{RJP2}
  \frac{\sqrt{3}\alpha_s}{8\pi}\langle 0\mid G\widetilde{G}\mid\eta \rangle=\langle 0\mid \partial_\mu J^{(0)}_{\mu 5}\mid \eta \rangle+\frac{1}{\sqrt{2}}\langle 0\mid \partial_\mu J^{(8)}_{\mu 5}\mid \eta \rangle,
  \ee
  \be \label{RJP3}
  \frac{\sqrt{3}\alpha_s}{8\pi}\langle 0\mid G\widetilde{G}\mid\eta' \rangle=\langle 0\mid \partial_\mu J^{(0)}_{\mu 5}\mid \eta' \rangle+\frac{1}{\sqrt{2}}\langle 0\mid \partial_\mu J^{(8)}_{\mu 5}\mid \eta' \rangle
  \ee

 Using Eq. (\ref{DivJ_FUMPhi}), (\ref{RJP1}), (\ref{RJP2}), (\ref{RJP3}) we deduce:

\be \label{RJP4Final}
R_{J/\Psi}=\left[\frac{m_{\eta'}^2}{m_\eta^2}\frac{f_0(-s_3s_8+c_3c_8c_0)+\frac{1}{\sqrt{2}}
f_8 (s_3c_8+c_3c_0s_8)}{f_0
(-c_3s_8-c_8c_0s_3)+\frac{1}{\sqrt{2}}f_8(c_8c_3-c_0s_3s_8)}\right]^2\left(\frac{p_{\eta'}}{p_{\eta}}\right)^3
\ee

Let us note that in obtaining this equation  only operator
relations for anomalies (\ref{RRPP1}, \ref{RRPP2}) were used
and one did not need to express the $\eta (\eta')$ mesons  fields
in the terms of divergencies of singlet  (and octet)  axial
currents.

\begin{figure}[t]
\begin{multicols}{2}
\hfill
\includegraphics[width=0.47\textwidth]{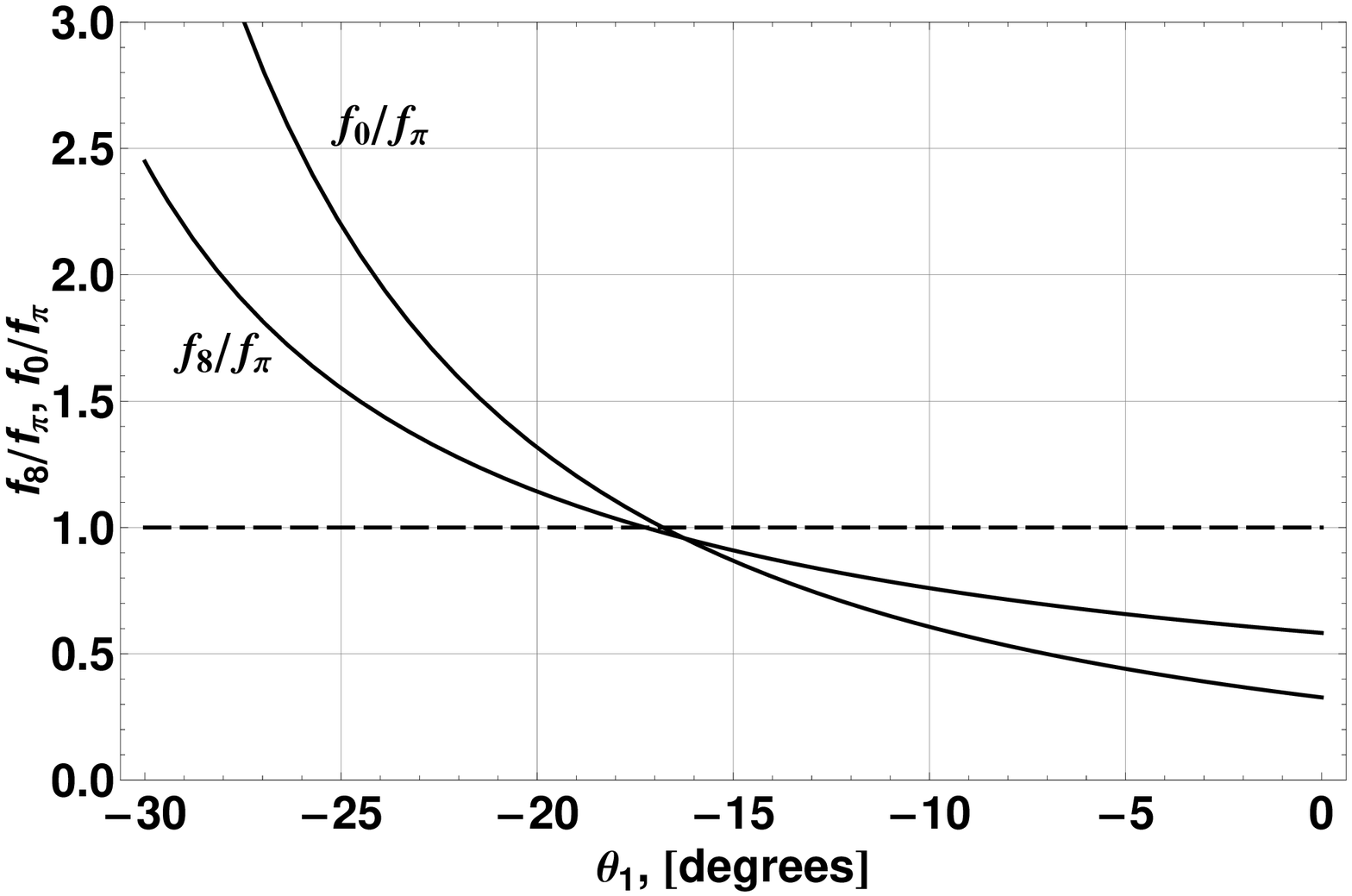}
\hfill \caption{Relationship between decay constants $f_8, f_0$
and mixing angle $\theta_1$ in the one-angle mixing scheme}
\label{1ang-plot2} \hfill
\includegraphics[width=0.47\textwidth]{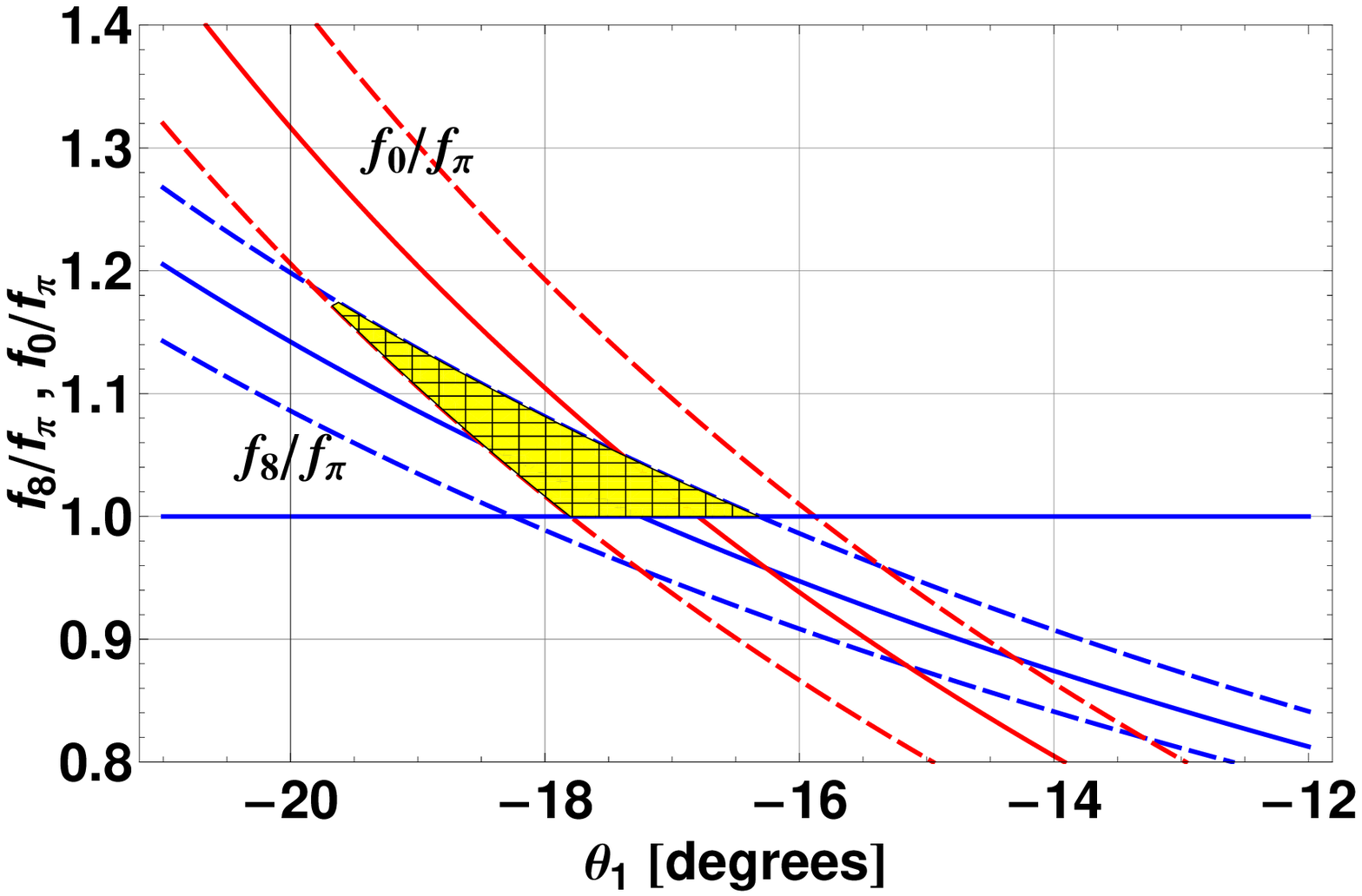}
\hfill \caption{Relationship between decay constants $f_8, f_0$
and mixing angle $\theta_1$ in the one-angle mixing scheme,
rescaled. Dashed curves denote experimental uncertainties, meshed
area indicates the range of parameters satisfying the condition
$f_8\gtrsim f_0\geq f_\pi $ } \label{1ang-plot3}
\end{multicols}
\end{figure}

If we use Eq.(\ref{RJP4Final}) for the case of one-angle mixing
scheme ($\theta_0=0$), for usual choice $f_8=1.28f_\pi$, $f_0=1.1f_\pi$ and corresponding angle value obtained  from
anomalous dispersive relation (\ref{J8_anom1})
$(\theta_1\equiv\theta_3+\theta_8=-22.1^\circ)$ we find
$R_{J/\Psi}=2.2$ which is in serious discrepancy with the
experimental value $R_{J/\Psi}=4.8 \pm 0.6$
\cite{Amsler:2008zzb}. Substituting the experimental value of
R into (\ref{RJP4Final}) and using anomaly equation
(\ref{J8_anom1}) one can get the dependencies  $f_8(\theta_1),
f_0(\theta_1)$ (Fig. \ref{1ang-plot2}).

As soon as we accept, that $f_8\gtrsim f_0\gtrsim f_\pi$ (for
different kind of justification see e.g. \cite{Feldmann:1998sh},
\cite{Leutwyler:1997yr}), it follows from Fig. \ref{1ang-plot2},
that the only possibility is $f_8\simeq f_0\simeq1$
$(\theta_1\simeq -18^\circ)$, which is quite far from the chiral
perturbation theory expectations \footnote{These values appear
to be preferable also in other mixing schemes as we will see
later.}. Taking into account all experimental errors (the dominant
contribution being provided by that of $R_{J/\Psi}$) one get Fig.
\ref{1ang-plot3} where the effects of these errors are indicated
by shaded areas. From this figure it is clear that the maximal
allowed value of $f_8$ is $f_8=1.2f_\pi(=f_0)$. Let us note that
these values correspond to minimal allowed value of
$R_{J/\Psi}=4.2$. Here the importance of more accurate
experimental value of $R_{J/\Psi}$ is already clear.

Let us pass to the more elaborated  schemes with more than one mixing angle which
were
offered in \cite{Leutwyler:1997yr,Feldmann:1998vh}.
In the most general scheme with 3 angles introduced in
Section 2 it is convenient to rewrite (\ref{RJP4Final}) in terms
of $\theta_1\equiv\theta_8+\theta_3,$
$\theta_2\equiv\theta_8-\theta_3$:

\be
\label{RJPsi5}R_{J/\Psi}=\left[\frac{m_{\eta'}^2}{m_\eta^2}\frac{f_0(c_1-c_2+c_0(c_1+c_2))+\frac{1}{\sqrt{2}}
f_8 (s_1-s_2+c_0(s_1+s_2))}{f_0
(-s_1-s_2-c_0(s_1-s_2))+\frac{1}{\sqrt{2}}f_8(c_1+c_2-c_0(c_2-c_1))}\right]^2\left(\frac{p_{\eta'}}{p_{\eta}}\right)^3
\ee The angle $\theta_1$ has an explicit physical sense. From the
definition of the  mixing matrix $\mathbf{U}$ (\ref{MixMat}) it
is obvious that the angle $\theta_1$ describes the overlap in the
$\eta-\eta'$ system with accuracy $\sim \theta_0^2/2$ and
coincides with their mixing angle at $\theta_0 \to 0$. It is
reasonable  to suppose that the glueball contribution to the
anomaly sum rule for non-singlet current (\ref{J8_anom_final}) is
rather small. This doesn't necessarily mean the extreme smallness
of $\eta-G$ mixing angle itself but rather the
cumulative effect of smallness of mixing angles and large mass
$m_G$. That's why we neglect the last term in the l.h.s. of
(\ref{J8_anom_final}) and rewrite it in terms of $\theta_1,
\theta_2 $: \be \label{J8_anom_newAng}
(c_1+c_2-c_0(c_2-c_1))+\beta(s_1-s_2+c_0(s_1+s_2))=2\xi. \ee

\begin{figure}[htp]
  \begin{center}
    \subfigure[$\theta_0=20^\circ$]{\label{fig3ang-a}\includegraphics[width=0.3\textwidth]{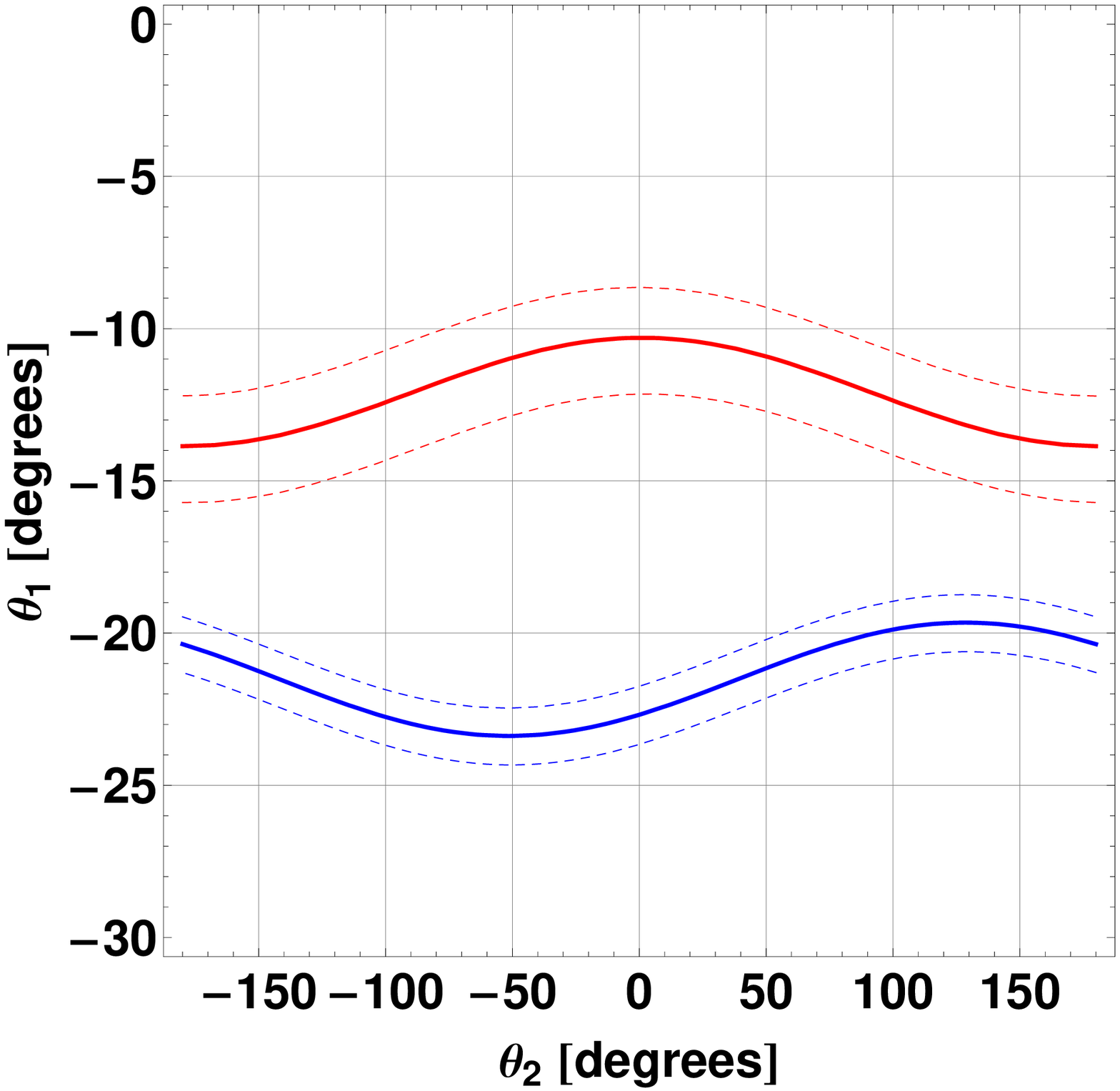}}
    \subfigure[$\theta_0=32^\circ$]{\label{fig3ang-b}\includegraphics[width=0.3\textwidth]{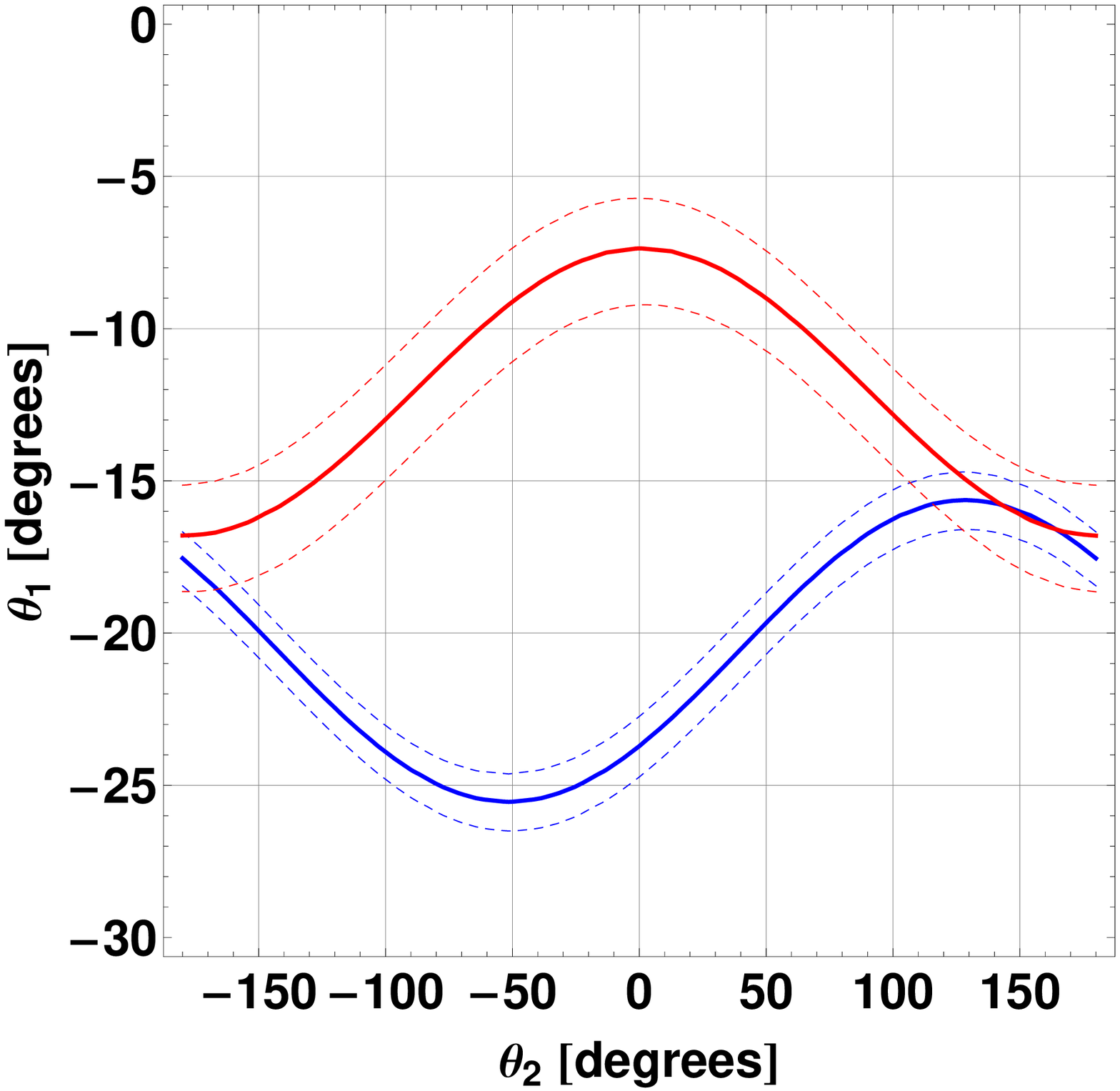}}
    \subfigure[$\theta_0=40^\circ$]{\label{fig3ang-c}\includegraphics[width=0.3\textwidth]{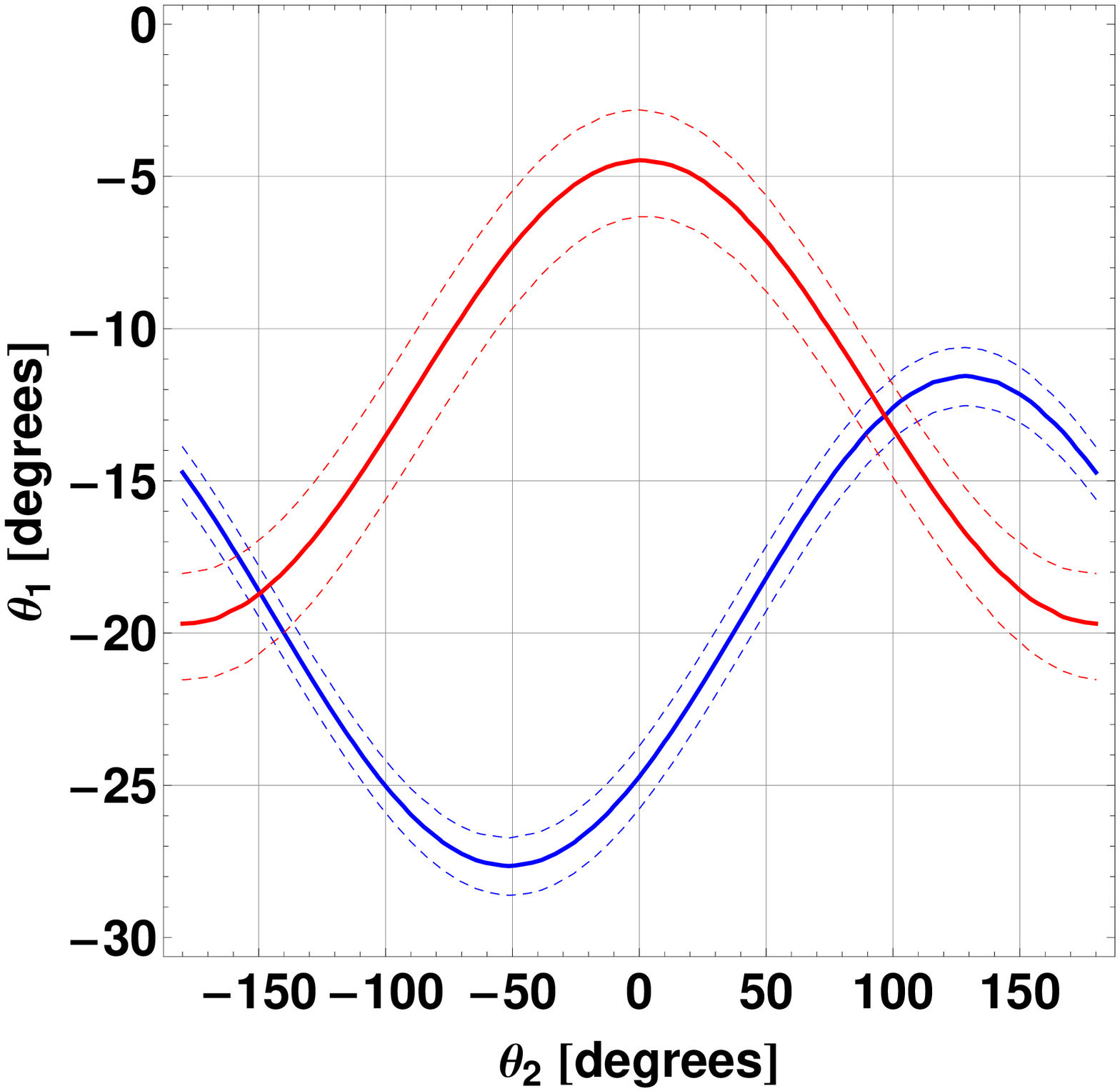}}
  \end{center}
  \caption{General (tree-angle) mixing scheme. Solutions of $R_{J/\Psi}$ equation (\ref{RJPsi5})(upper, red online curves) and anomaly condition (\ref{J8_anom_newAng}) (lower, blue online curves) for $f_8=1.28f_\pi, f_0=1.1f_\pi$ and different choice of $\theta_0$. The dashed curves indicate experimental uncertainties.}
  \label{fig3ang}
\end{figure}

The solutions of (\ref{RJPsi5}) (upper curves, red online),
and (\ref{J8_anom_newAng}) (lower curves, blue online) are shown on
Fig. \ref{fig3ang} for customary choice of decay constants
$f_8=1.28f_\pi, f_0=1.1f_\pi$  and three different angles
$20^\circ, 32^\circ, 40^\circ$. We see from these figures that at
$\theta_0=20^\circ$ there is no intersection. The first common
solution appears at larger angle $\theta_0=32^\circ$ when the
curves touch each other. As $\theta_0$ grows, at
$\theta_0>32^\circ$ two different solutions appear. While for
these two solutions $\theta_2$ are significantly different,
the solutions for $\theta_1$ are limited to relatively
narrow region $\theta_1=-13^\circ\div-18^\circ$. This is not
surprising because of the physical sense of $\theta_1$ mentioned above \footnote{Other solutions
not shown on the figures
appear only at large
$\theta_0 > 70^\circ$ which clearly has no physical sense.}.
Taking into account the experimental uncertainties we see that
the minimal allowed value of $\theta_0$ slightly decreases to
$27^\circ$.

All this numerical analysis showed that for
$f_8=1.28,f_0=1.1$ we need a substantial glueball admixture
$\theta_0>27^\circ$. The minimal value of $\theta_0$ decreases
only if $f_8$ decreases. The results of analysis for different
$f_8$ and $f_0$ are presented in Table \ref{Table},
\begin{table}
\begin{tabular}{|c|c|c|c|}\hline
\backslashbox { $f_0/f_\pi$ }{ $f_8/f_\pi$ } & 1.28 & 1.1 & 1.0
\\ \hline 1.28& $24^{+5}_{-6}$ & &  \\\hline 1.1 &
$32^{+3.5}_{-4.5}$ &$17.5^{+6}_{-12}$  & \\\hline 1.0 &
$35^{+2.5}_{-3.5}$ & $23.5^{+5}_{-6.5}$ & $10^{+9}_{-10}$ \\\hline
\end{tabular}
\caption{Tree-angle mixing scheme. Minimal possible mixing angle
$\theta_0$ (in degrees) for different values of decay constants
$f_0$ and $f_8$} \label{Table}
\end{table}
from which one can make an important  conclusion that relatively
small glueball admixture even within experimental uncertainties
is possible only for $f_8=f_0$ (and most probably
$f_8=f_0=f_\pi$.)

\section{Two-angle mixing schemes}
Up to a moment our analysis was quite general. It is instructive
to consider some particular cases which are currently discussed
in literature. Starting from our general scheme, one can easily
perform a reduction to partial mixing schemes with 2 angles. Here we have 2
distinct cases when in  (\ref{MixMat})

i) $\theta_3=0$
or

ii)  $\theta_8=0$.

 \textbf{1. The case} {$\mathbf{\theta_3=0}$.

This case was introduced in \cite{Kou:1999tt}, and was widely
used by KLOE collaboration in a set of recent papers
\cite{Ambrosino:2006gk}-\cite{Ambrosino:2009sc}. \footnote{Note
that in these papers authors use quark basis, see (\ref{qbasis}).
The connection between  notations of mixing angles in
\cite{Ambrosino:2009sc} and our notations: $\theta_0=-\phi_G$,
$\theta_8=\phi_P-\alpha$, where $ \alpha= \arctan\sqrt{2/3}
\simeq 54.7^\circ$.} This choice clearly means that there is no
mixing of $\eta$ with additional scalar state noted in
(\ref{Phi}) as $g$, i.e. $\eta$ is a mixture of $\phi_8$ and
$\phi_0$ only. In this case our equations (\ref{J8_anom_final}),
(\ref{RJP4Final}) are simplified as follows: \be
\label{J8anomKLOE} c_8+\beta c_0s_8+\gamma s_0s_8=\xi, \ee

\be \label{RJP5KLOE}
R_{J/\Psi}=\left[\frac{m_{\eta'}^2}{m_\eta^2}\frac{f_0(c_8c_0)+\frac{1}{\sqrt{2}}
f_8 (c_0s_8)}{f_0
(-s_8)+\frac{1}{\sqrt{2}}f_8(c_8)}\right]^2\left(\frac{p_{\eta'}}{p_{\eta}}\right)^3
\ee
Note that $\theta_8$ in this scheme defines $\eta-\eta'$
mixing, as one can easily see from Eq. (12). In further analysis
we suppose that $\gamma$ cannot exceed 2 for any reasonable
values of $\Gamma_{G \to 2\gamma}$ (i.e.  $\Gamma_{G\to
2\gamma}/m_G^3\lesssim 4\Gamma_{\eta\to2\gamma}/m_{\eta}^3)$.
This restriction corresponds to the assumption  that 2-photon
decay widthes of pseudoscalar mesons grow like the third power of
their masses, or in other words  the glueball coupling to quarks
to be of the same order as for the meson octet
states.\footnote{If it grows with mass according to the
na\"{\i}ve dimensional arguments, the glueball width will be even
smaller. We will discuss the relation between coupling constants
and decay widthes later.}

The results of numerical analysis are shown on  Fig.
\ref{figKLOE}. These figures show the dependence of the glueball
contribution $\gamma$ on the angle
$\theta_8$ for different values of $f_8$ and $f_0$.  

On Fig. \ref{figKLOEa} \footnote{We limit ourselves to negative
$\theta_8$ since this mixing angle in this scheme clearly have a
sense of $\eta-\eta'$ mixing angle.} the dependence $\gamma
(\theta_8)$ is shown for $f_8=1.28f_\pi$ and
$f_0=(1.1,1.28)f_\pi$. The dotted lines corresponds to
experimental uncertainties, the uncertainty for R  being dominant.
From this figure one can see that for any $f_0<f_8=1.28f_\pi$  we
get $\gamma>2$. One can achieve $\gamma\simeq2$ only for
$f_0\simeq1.28\simeq f_8$ (at the lower value of $R_{J/\Psi}=4.2$
). The corresponding values of mixing angles are $\theta_8=-(14
\div 17)^\circ$, $\theta_0= (12 \div 30)^\circ$ .

On the  Fig. \ref{figKLOEb} the case  $f_8=1.1$ for two choices
of $f_0=(1.1,1.0)$ is shown. Again, the reasonable values
$\gamma\lesssim2$ is achievable only for $f_0\simeq1.1f_\pi\simeq
f_8$. The corresponding values of mixing parameters are
$\theta_8=-(12 \div 18)^\circ$, $\theta_0= (5 \div 35)^\circ$

And, at last, consider the choice $f_0=f_8=1.0f_\pi$ shown on
Fig. \ref{figKLOEc}. The corresponding values of mixing
parameters are $\theta_8=-(10 \div 18)^\circ$, $\theta_0= (5 \div
37)^\circ$ Note that in this case the region of relatively small
$\gamma\lesssim 1$ and $\theta_0 \sim 5$  are achievable.

Let us notice that for $f_8>1.28f_\pi$  the minimal value of
$\gamma$ is growing, e.g. for large $N_c$ value $f_8=1.34f_\pi$
(and any $f_0$), $\gamma \gtrsim 2$ within experimental errors of $R_{J/\Psi}$.

We can conclude, that in this scheme the glueball admixture is
bounded from below to minimal value $\theta_0>5^\circ$ for any
$f_8\gtrsim f_0\gtrsim f_\pi$. The relatively small glueball
admixture $\theta_0\lesssim 10^\circ$ is possible only for
$f_8\simeq f_0\simeq f_\pi$.

Let us stress, that here (like in general 3-angle mixing scheme)
the values $f_8\simeq f_0$ is much more preferable. All these
results directly follow from the specific choice of mixing
scheme, axial anomaly condition and $R_{J/\Psi}$.

It is instructive to compare our results with those obtained in
\cite{Ambrosino:2009sc}, where the analysis of
$BR(\phi\to\eta'\gamma)/BR(\phi\to\eta\gamma)$ together with
several vector meson radiative decays to pseudoscalars was
performed in the same mixing scheme.

If we substitute their values of $\theta_8=(-14.5\pm 1),
\theta_0=7.5^\circ$ to our dispersive relation and take into
account possible sources of inaccuracy  we can see that we
directly get the dependence $\gamma(f_8)$ shown on Fig.
\ref{figK2}. The large slope is due to small mixing angles
$\theta_0$ and $\theta_8$.

\begin{figure}[htp]
  \begin{center}
    \subfigure[$f_8=1.28f_\pi$]{\label{figKLOEa}\includegraphics[width=0.3\textwidth]{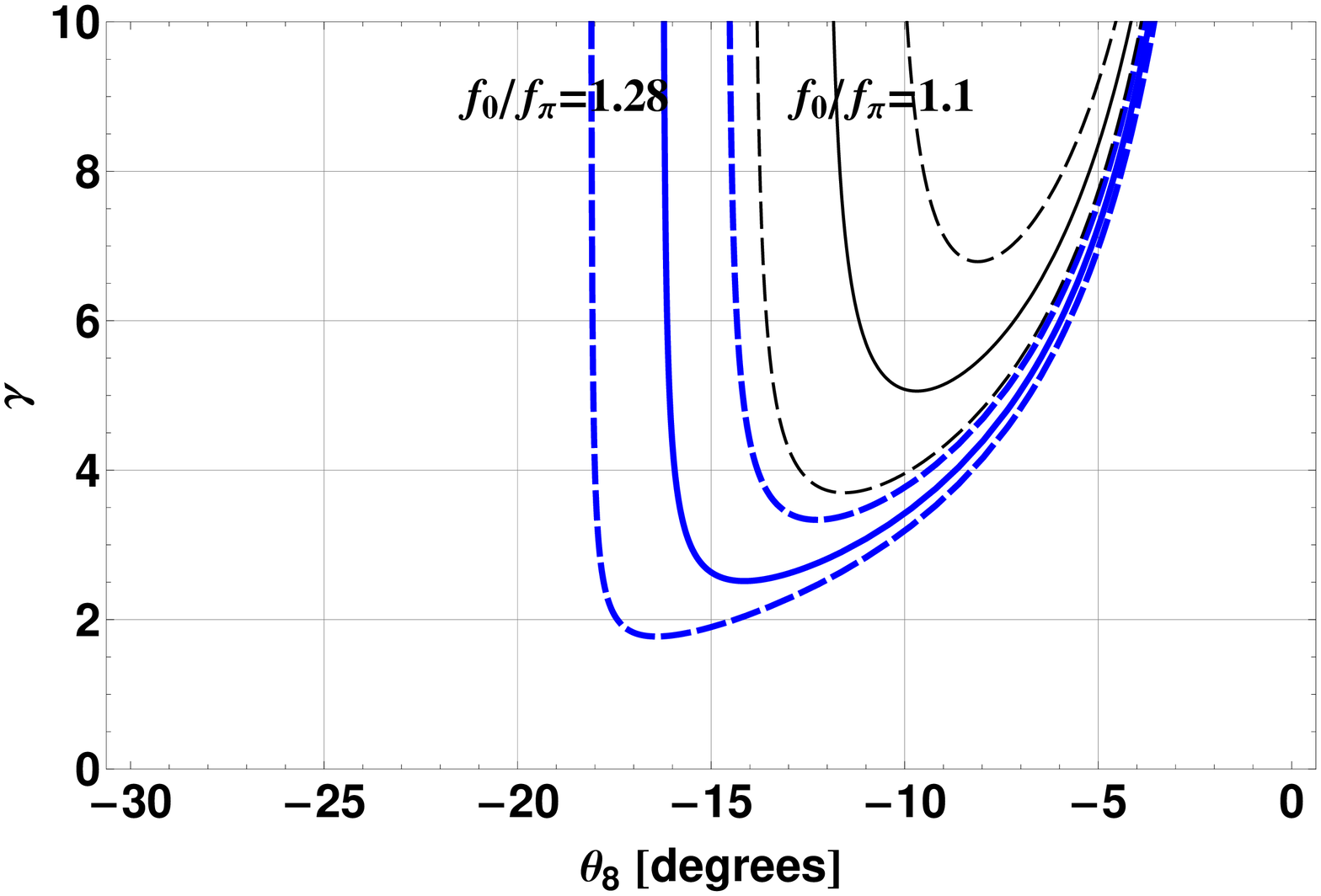}}
    \subfigure[$f_8=1.1f_\pi$]{\label{figKLOEb}\includegraphics[width=0.3\textwidth]{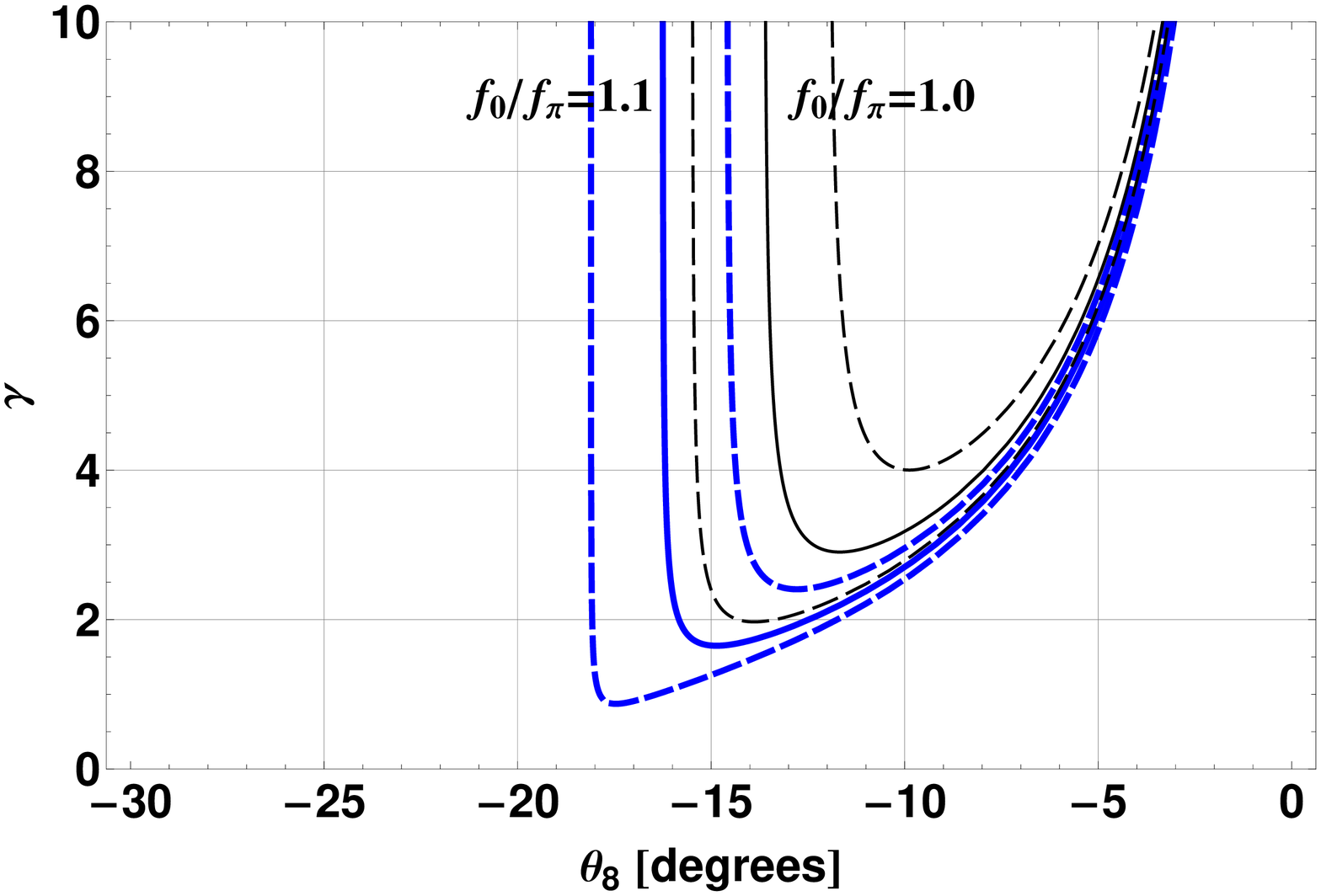}}
    \subfigure[ $f_8=1.0f_\pi$]{\label{figKLOEc}\includegraphics[width=0.3\textwidth]{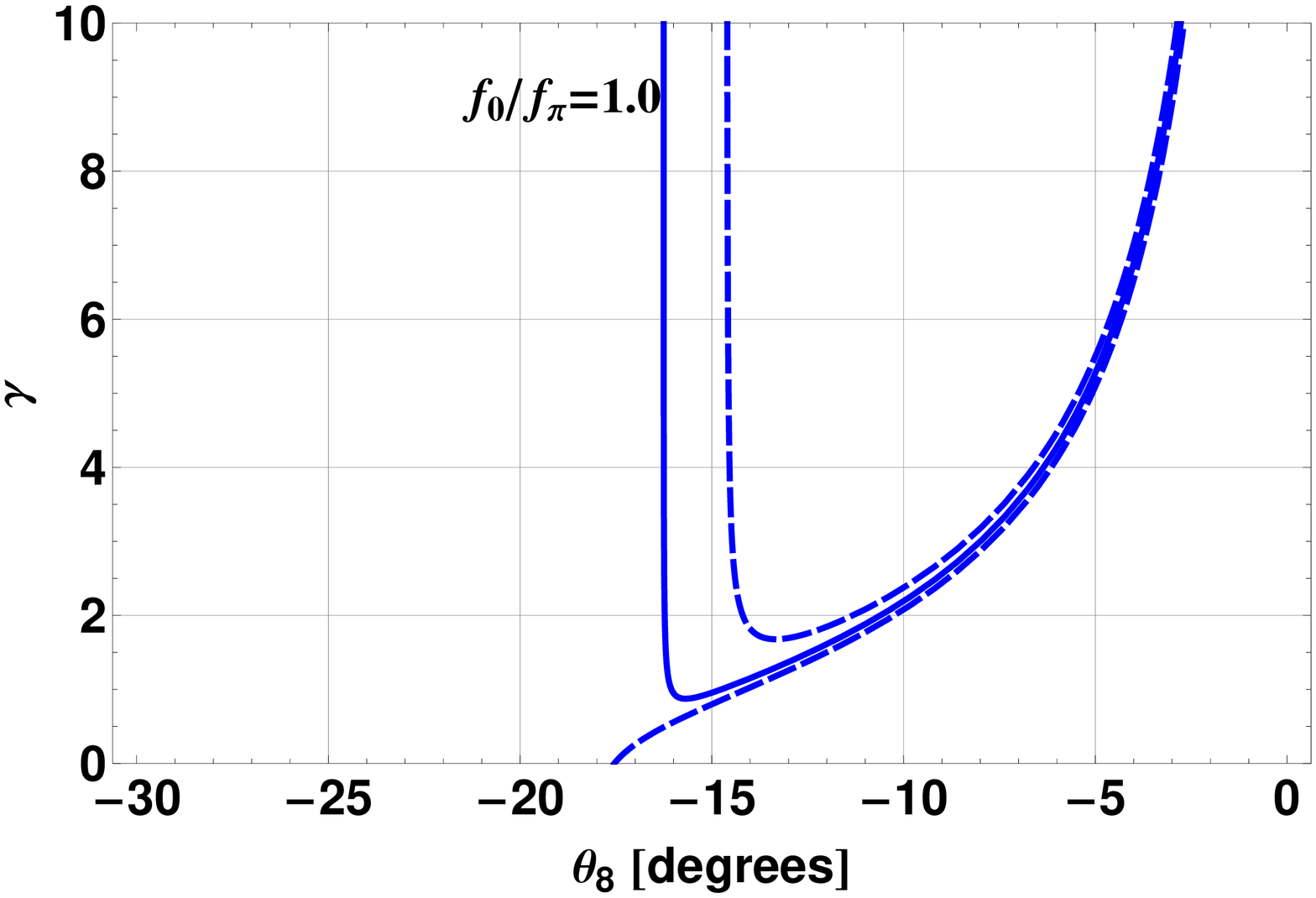}}
  \end{center}
  \caption{Glueball contribution parameter $\gamma$ as a function of mixing angle $\theta_8$ for different decay constants.
  Dashed curves indicate errors arising from experimental data.}
  \label{figKLOE}
\end{figure}

One can see that the reasonable value $\gamma<2$ corresponds to
$f_8\sim1 ((1.00-1.05)f_{\pi})$. Thus anomaly constraint
contradicts the preferred value \cite{Leutwyler:1997yr}
$f_s=1.34f_\pi$ $(f_8=1.28f_\pi)$ which is used in
\cite{Ambrosino:2009sc}. However, the constant $f_s$ enters only
the expression for radiative decay width $\eta'\to
2\gamma$ (Eq.1.4 in \cite{Ambrosino:2009sc}). The change of $f_s$
from $f_s=1.34f_\pi (f_8=1.28 f_\pi)$  to the value $f_s=f_\pi (f_8=f_\pi)$ in this equation may be compensated by the change of
$\Psi_G$ within the claimed accuracy. Therefore at the current
level of accuracy the analysis of KLOE is compatible with ours.
 The  combination of the KLOE results  with both our constrains
 (i.e. anomaly  and $R_{J/\Psi}$, (\ref{J8anomKLOE}, \ref{RJP5KLOE}) leads
 to conclusion that
 $f_8=f_0=f_\pi$ which is in agreement with the analysis presented above. \\

\begin{figure}[t]
\begin{multicols}{2}
\hfill
\includegraphics[width=0.47\textwidth]{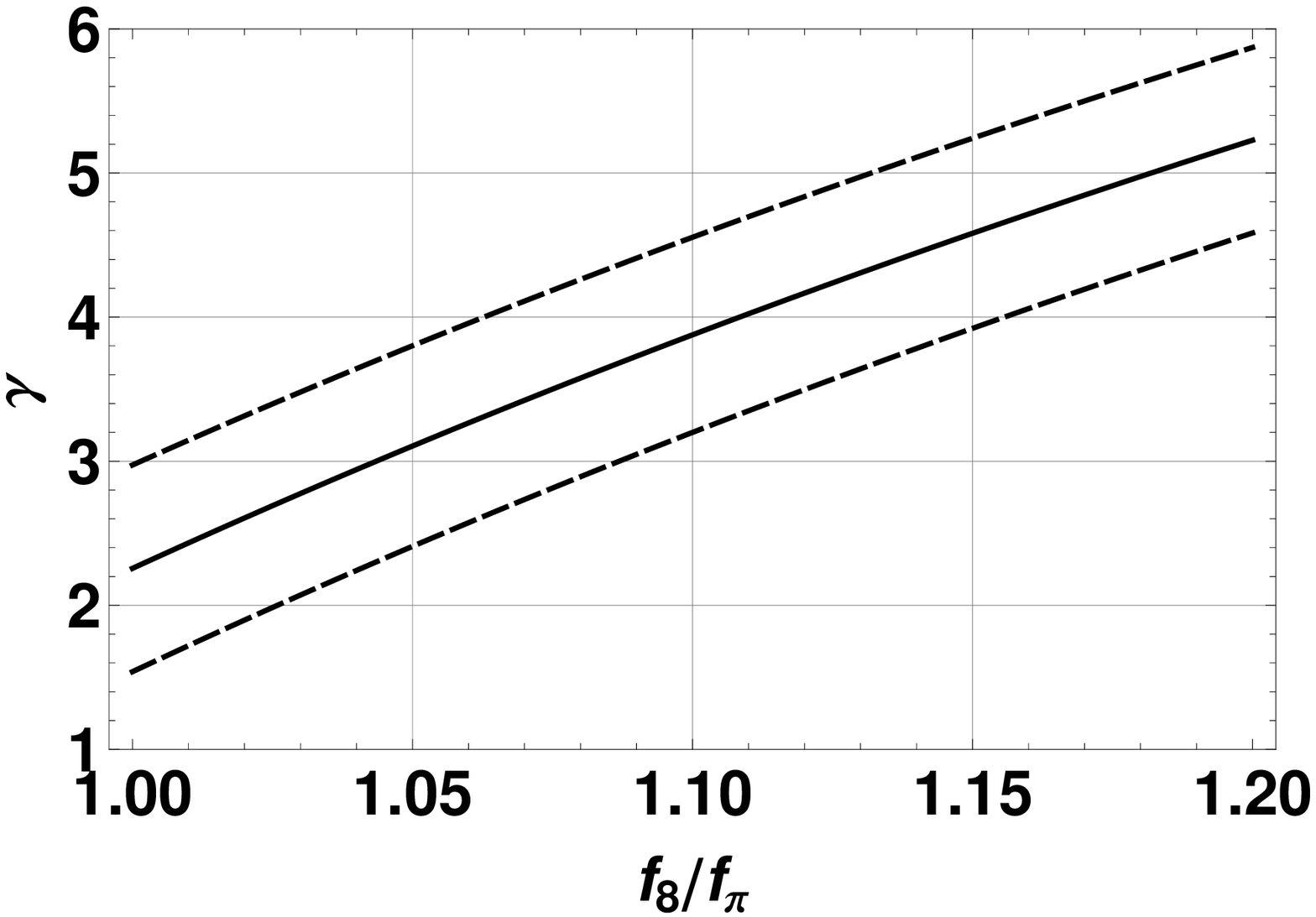}
\hfill \caption{Two-angle mixing scheme, case $\theta_3=0$.
Glueball contribution parameter $\gamma$  as a function of decay
constant $f_8$ (in units of $f_\pi$) for
$\theta_8=~(-14.5\pm 1)^\circ$, $\theta_0=7.5^\circ$}
\label{figK2} \hfill
\includegraphics[width=0.47\textwidth]{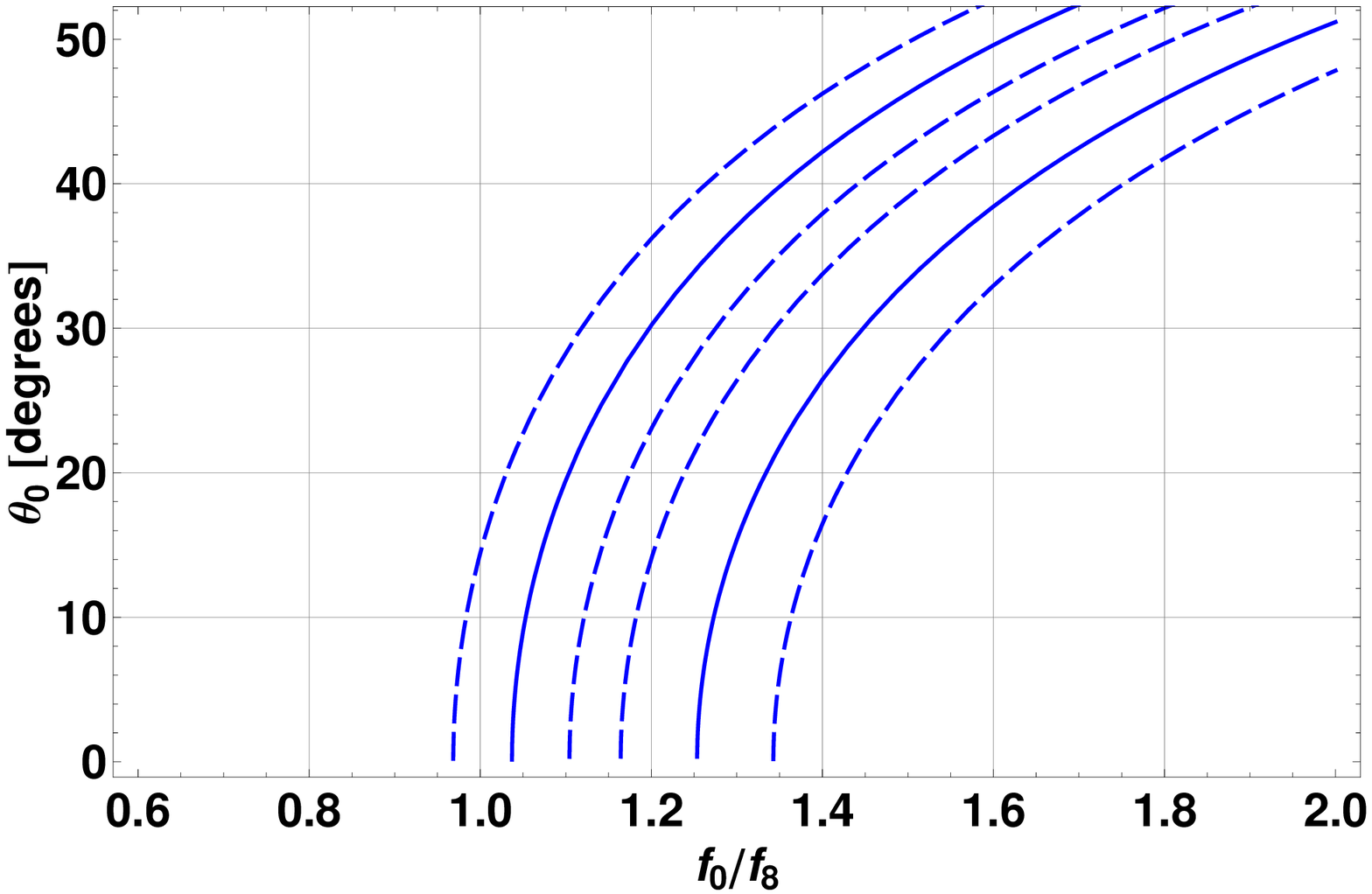}
\hfill \caption{Two-angle mixing scheme, case $\theta_8=0$. The
dependence of the mixing angle $\theta_0$ on the ratio of decay
constants $f_0/f_8$ together with experimental uncertainties.}
\label{figC}
\end{multicols}
\end{figure}

 \textbf{2. The case} {$\mathbf{\theta_8=0}$.

One can consider another particular case of general mixing
supposing $\theta_8=0$ in (\ref{MixMat}). This kind of a
two-angle mixing scheme was used recently in \cite{Cheng:2008ss}.
This scheme implies no glueball coupling to $J_{\mu5}^{8}$, which
means that the glueball contribution to anomaly relation is {\it exactly} zero.
Therefore, the anomaly relation  is exactly the same as for the
one-angle mixing scheme so the dependence $f_8(\theta_3)$ is the
same and shown on Fig. \ref{1ang-plot1}. Recall,  that this
relation is precise due to reasons, mentioned in Section 3.

The relation for $J/\Psi$ radiative decay ratio for this case can
be directly obtained from (\ref{RJP4Final}) with $\theta_8=0$.
This relation together with axial anomaly relation allow us to
get the dependence of the ratio of decay constants $f_0/f_8$ as a
function of $\theta_0$ (see Fig. \ref{figC}). \footnote{ As it was
noted in Section 4 the ratio $f_0/f_8$ should not exceed 1.}
Stripes bounded by dashed lines show the effects of experimental
uncertainties. From Fig. \ref{figC} it is clear that the solution at
$f_8=1.28f_\pi$ contradicts to the condition $f_0/f_8<1$
discussed above for any $\theta_0$. The only solution compatible
with $f_0/f_8\leq1$ is the one at $f_8=(1-1.05)f_\pi$ (the left
stripe), the corresponding
$\theta_0=0^\circ \div 14^\circ$, while the angle $\theta_3$
(having the sense of the $\eta-\eta'$ overlap)
can vary in a very
narrow region $\theta_3=-17^\circ \div-18^\circ$.

So one can conclude that this scheme even within current
experimental uncertainties demands $f_8=f_0=f_\pi$ and
small glueball admixture $\theta_0<14^\circ$.

Let us now examine the numerical results of \cite{Cheng:2008ss}.
Implementation of our procedure with their angle $\phi=42.4^\circ$ ($\theta_3=-12.3$) leads
to $f_8=0.8$ and contradicts to $f_8/f_\pi>1.0$ being the robust expectation
from the ChPT.

As soon as the accuracy of the extracted \cite{Cheng:2008ss} parameters is not
available at the moment it is difficult to conclude whether this disagreement
is statistically significant.

\section{Summary}

In this paper we presented the detailed analysis of the role of
Abelian axial anomaly in the mixing of both light and heavy
pseudoscalar states. We found that the anomaly imposes the severe
constraints for the meson couplings and mixing angles. There is
also a delicate interplay between pseudoscalar state, which we
call glueball without specifying its nature, and light mesons.

We offered a (new, to our best knowledge) rigorous approach in
$SU(3)$ basis to the consideration of mixing of pseudoscalar
mesons. Our approach is quite similar to that for the mixing of
massive neutrino, where the number of states with a definite mass
may exceed  the number of flavor states. In this sense the
appearance of extra singlet mesons is similar to the role of
sterile neutrinos.

We use the dispersive representation of axial anomaly for
$J_{\mu5}^8$. The advantage of such representation is the high
and controlled accuracy due to suppression of higher resonances
and its independence on the quark masses. The use of $SU(3)$
basis allows us to limit ourself to more theoretically clear case
of non-singlet current receiving contribution only from Abelian
anomaly. Let us note, that our equation (\ref{J8_anom1}) may be
formally derived in the other pioneering approaches using
standard local anomaly relations (see e.g. \cite{Ball:1995zv} in
the one-angle mixing scheme.

As a supplementary input we use experimental data for the ratio
of radiative decays $R_{J/\Psi}\equiv(\Gamma(J/\Psi)\to
\eta'\gamma)/(\Gamma(J/\Psi)\to \eta\gamma)$ which is
theoretically safe and provides additional restriction.

We found that for the one-angle mixing scheme the only
reasonable  solution is $f_8\simeq f_0 = f_\pi,  \theta_1=-18^\circ$.

We proved that any scheme with more than one angle unavoidably
demands additional singlet admixture and  considered the general
scheme with 3 angles with one lowest additional singlet state,
which we denoted as a glueball without any specification of its
nature.

It was found, that the combination of our two inputs (anomaly
relation and $R_J\Psi$ data) leads to a rather strict
constraints. The values $f_8=1.28f_\pi$ inevitably leads to a
large  glueball admixture $\theta_0\simeq 30^\circ$. At the same
time, the reasonably small glueball admixture $\theta_0<10^\circ$
is possible only for $f_0 \simeq f_8 = 1.0 f_\pi$, which is far
from expectations, based on chiral  perturbation theory.

We checked this general observation by considering some
particular cases, when the general mixing matrix (with  3 angles)
is reduced to the $\eta-\eta'-G$ mixing scheme with two angles.

The first one is the
mixing scheme with $\theta_3=0$ which was investigated in a set of
recent papers by  KLOE collaboration. We concluded that
application of our approach to this scheme unavoidably leads to
$f_8/f_0\approx1$ for any reasonable values of glueball
two-photon decay width. Moreover, combining our constraints with
the angles $\theta_8 = 14.5^\circ, \theta_0 = 7.5^\circ$
obtained in KLOE analysis based on the additional set of decays,
we again found that $f_8\approx f_0= f_\pi$. These values are
still consistent with the results of their fits within their
accuracy.

We also consider another two-angle mixing scheme where
$\theta_8=0$. In this case the only possible solution compatible
with constraint $f_8\geq f_0$ is $f_8=f_0=f_\pi$ with high
accuracy.

As a result, in all the considered schemes the relation
$f_8\simeq f_0$( $\simeq f_\pi$ most likely) holds. This marks a
sort of new manifestation of $SU(3)$ symmetry
and at the same
time the possible violation of chiral perturbation theory
expectations.
The possible origin of such a symmetry pattern may be the smallness
of the strange quark mass (squared) with respect to nucleon one
while it is still much larger than (genuine) higher twist parameters
and may be treated sometimes as a heavy one \cite{Polyakov:1998rb}.

The significant progress may be achieved by the more accurate
determination of $R_{J/\Psi}$, in particular, at BES-III
accelerator, complementing its vast program (see \cite{Asner:2008nq}, Section 17.2.2).
While our conclusion $f_8\simeq f_0$ is more robust
and is valid for all values of $R_{J/\Psi}$ within current
experimental limits, the stronger result $f_8=f_\pi$ may be
questioned by the more accurate data. Therefore these
measurements will provide a new test of chiral perturbation
theory predictions.

Although our approach already provided the important new
constraint for the analysis in pseudoscalar channel, the global
fit exploiting the dispersive representation of axial anomaly
remains to be done.

\textbf{Acknowledgements.}

The authors are indebted to D.~I.~Diakonov, J.~Horejsi,
B.~L.~Ioffe, A.~V.~Kisselev, P.~Kroll, V.A.~Naumov, T.~N.~Pham,
M.~V.~Polyakov and C.~D.~Roberts for useful discussions and
correspondence. O.V.T. is also thankful to
D.V.~Dedovich, G.A.~Shelkov and  A.S.~Zhemchugov for the discussions of
the physical program of BES-III.
This work was supported in part by RFBR   (Grants
09-02-00732, 09-02-01149), by the funds from EC to the project
"Study of the Strong Interacting Matter" under contract N0.
R113-CT-2004-506078 and by  CRDF  PROJECT  RUP2-2961-MO-09.

\end{document}